\documentclass[preprint]{jpsj3}
\usepackage{color}

\title{Exact Realization of a Quantum-Dimer Model in Heisenberg Antiferromagnets on a Diamond-Like Decorated Lattice}
\author{Yuhei Hirose, Akihide Oguchi, and Yoshiyuki Fukumoto}
\inst{Department of Physics, Faculty of Science and Technology,\\ Tokyo University of Science, Noda, Chiba 278-8510, Japan}
\recdate{\today}
\abst{
We study Heisenberg antiferromagnets on a diamond-like decorated square lattice perturbed by further neighbor couplings.
The second-order effective Hamiltonian is calculated and 
the resultant Hamiltonian is found to be a square-lattice quantum-dimer model with a finite hopping amplitude and no repulsion,
which suggests the stabilization of the plaquette phase.
Our recipe for constructing quantum-dimer models can be adopted for other lattices 
and provides a route for the experimental realization of quantum-dimer models.
}

\begin{document}
\maketitle

%%%%%%%%%%%%%%%%%%%%%%%%%%%%%%%%%%%%%%%%%%%%%%%%%%%%%%%%%%%%%%%%%%%%%%%%%%%%%%%%%%%%%%
\section{Introduction}\label{sec:1}
%%%%%%%%%%%%%%%%%%%%%%%%%%%%%%%%%%%%%%%%%%%%%%%%%%%%%%%%%%%%%%%%%%%%%%%%%%%%%%%%%%%%%%

The exploration of resonating valence bond (RVB) states is one of the central issues in condensed matter physics\cite{Anderson1973}.
The simplest effective model is the Rokhsar$-$Kivelson quantum-dimer model (QDM)\cite{Rokhsar1988},
\begin{equation}
\begin{minipage}[c]{.92\textwidth}
\begin{center}
\includegraphics[width=1\linewidth]{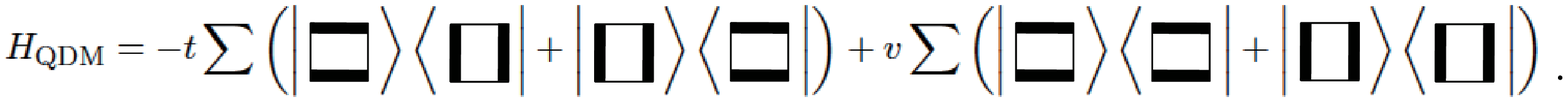}
\end{center}
\end{minipage}
\end{equation}
It is naively expected that the pair hopping leads to phenomena stemming from the resonance of various dimer-covering states, and 
the interaction tends to stabilize the crystallization of dimers.

Much theoretical effort has been devoted to clarifying ground-state phase diagrams as a function of $v/t$ for QDMs on various lattices.
When pair hopping is dominant, i.e., when $|v/t|\ll 1$, fascinating phases have been found: the plaquette phase for a square lattice,
the $\mathbb{Z}_2$ RVB liquid for nonbipartite lattices, the $U(1)$ RVB liquid for a simple cubic lattice, and so forth\cite{Leung1996,Moessner2008}.
However, the issue of the experimental realization of QDMs is unclear. 
The QDM may be an effective theory for the $S=1/2$ kagome-lattice Heisenberg antiferromagnet\cite{Yan2011,Depenbrock2012}
but its mathematical interpretation is intractable.

Actually, a great deal of effort has been devoted to constructing QDMs from the original quantum spin Hamiltonian. However, existing construction methods for QDMs are extremely complicated because these methods invoke multiple spin interactions\cite{Fujimoto}.  It is not yet clear how to derive QDMs from the simple Heisenberg model.

Very recently, Morita and Shibata have proposed a method for constructing the QDM Hilbert space\cite{Morita2016}.
In their method, a lattice is considered, and all the edges of the lattice are replaced with a diamond unit.
The model Hamiltonian contains only  Heisenberg-type exchange terms, which is very important for experimental realizations, as an example of a quantum simulator using optical lattices, \cite{Bloch2008} and the ground-state manifold is equivalent to the dimer covering of the original lattice.
For example, the above-mentioned replacement gives a diamond chain when we start with a linear chain\cite{Takano1996}.
For a range of coupling parameters, the ground states are known to be twofold degenerate tetramer-dimer states\cite{Takano1996}.
When we regard a tetramer as a ``dimer'' in a QDM, the tetramer-dimer states are identical to the dimer-covering states of the linear chain.
In a similar way, the replacement gives a diamond-like decorated square lattice\cite{Canova2010} when starting with a square lattice.
For a suitable range of coupling parameters, we have square-lattice dimer-covering states as the ground-state manifold.  

In this paper, we consider Heisenberg antiferromagnets on a diamond-like decorated square lattice, which is somewhat different from the situation in Ref.~11,
and introduce a perturbation that breaks the degeneracy of the ground-state manifold.
We calculate the second-order effective Hamiltonian and show that it is exactly the same as the square-lattice QDM.
Therefore, we present a construction method for a QDM by simply using the simple Heisenberg coupling.
It will be shown that the QDM has $v=0$; thus, the resonance effect determines a unique ground state.

This paper is organized as follows.
The Heisenberg model on a diamond-like decorated square lattice and the perturbation Hamiltonian are defined in sect. 2.
In sect. 3, we derive a square-lattice QDM as a second-order effective Hamiltonian.
In sect. 4, we summarize the results obtained in this study.

%%%%%%%%%%%%%%%%%%%%%%%%%%%%%%%%%%%%%%%%%%%%%%%%%%%%%%%%%%%%%%%%%%%%%%%%%%%%%%%%%%%%%%
\section{Model}\label{sec:2}
%%%%%%%%%%%%%%%%%%%%%%%%%%%%%%%%%%%%%%%%%%%%%%%%%%%%%%%%%%%%%%%%%%%%%%%%%%%%%%%%%%%%%%

\subsection{Diamond unit and tetramer ground state}

\begin{figure}[t]
\begin{center}
\includegraphics[width=.6\linewidth]{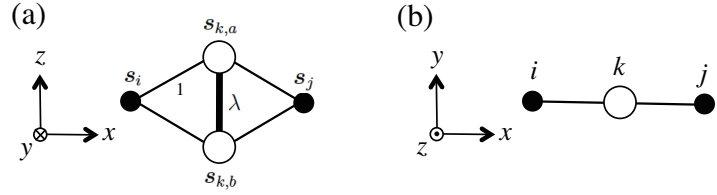}
\end{center}
\caption{(a) Top and (b) side views of the diamond unit.
The position of the site $k$ is given by $\mib{r}_k=\frac{1}{2}(\mib{r}_i+\mib{r}_j)$.}
\label{fig:1}
\end{figure}

For the diamond unit in Fig.~\ref{fig:1}, we introduce the site indices shown in this figure and define
\begin{equation}
   h_{i,j}=(\mib{s}_i+\mib{s}_j)\cdot(\mib{s}_{k,a}+\mib{s}_{k,b})+\lambda \left(\mib{s}_{k,a}\cdot\mib{s}_{k,b}+\frac{3}{4}\right),
\end{equation}
where the magnitude of the spin operators is $1/2$. 
We call $\mib{s}_i$ and $\mib{s}_j$ the edge spins and the pair $(\mib{s}_{k,a},\mib{s}_{k,b})$ a bond spin-pair.
The most important property of $h_{i,j}$ is that the total spin of the bond spin-pair is a good quantum number\cite{Takano1996}.

%----------------------------------------------------
%----------------------------------------------------

When the coupling constant for a bond spin-pair is less than 2, $\lambda<2$, the lowest eigenvalue of $h_{i,j}$ is $-(2-\lambda)$,
and the corresponding eigenvector $|\phi^g\rangle_{i,j,k}$ (tetramer ground state) is given by
\begin{equation}
   |\phi^g\rangle_{i,j,k}=\frac{1}{\sqrt{3}}
   \left(
   |\!\uparrow\uparrow\rangle_{i,j}|t^-\rangle_k+|\!\downarrow\downarrow\rangle_{i,j}|t^+\rangle_k
   -\frac{|\!\uparrow\downarrow\rangle_{i,j}+|\!\downarrow\uparrow\rangle_{i,j}}{\sqrt{2}}|t^0\rangle_k
   \right),
\label{eq:2} 
\end{equation}
where $\{|t^+\rangle, |t^0\rangle, |t^-\rangle\}$ represents the triplet states of a bond spin-pair.
Note that the bond spin-pair is in triplet states but the magnetism is screened by the edge spins; thus, $\phi^g$ is nonmagnetic and
\begin{equation}
   (\mib{s}_i+\mib{s}_j+\mib{s}_{k,a}+\mib{s}_{k,b})^2|\phi^g\rangle_{i,j,k}=0.
\end{equation}
On the other hand, the eigenfunctions of $h_{i,j}$ are simple product states when the total spin of the bond spin-pair is zero:
\begin{equation}
   |\sigma,\sigma'\rangle_{i,j}|s\rangle_k,
\label{eq:5} 
\end{equation}
where $\sigma,\sigma'=\uparrow$ or $\downarrow$ and $|s\rangle_k$ represent the singlet state of the bond spin-pair.
Equation (\ref{eq:5}) indicates that bond spin-pairs with $S^{\rm{tot}}=0$ in a connected system of diamond units segment the system into clusters constructed by the spin-pairs with $S^{\rm{tot}}=1$ and edge spins.
This concept is frequently used in the following study of a diamond-like decorated square lattice.

\subsection{Hamiltonian}%%%%%%%%%%%%%%%%%%%

\begin{figure}[t]
\begin{center}
\includegraphics[width=.70\linewidth]{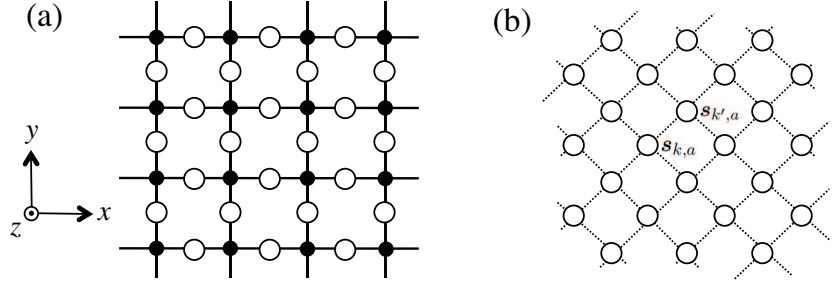}
\end{center}
\caption{(a) Structure of the diamond-like decorated square lattice,
where the direction of the bond spin-pairs is along the $z$-axis.
The edge spins are in the plane $z=0$.
There is a plane composed of $\{\mib{s}_{k,a}\}$ in the region $z>0$
and another plane composed of $\{\mib{s}_{k,b}\}$ in the region $z<0$.
(b) Schematic representation of the perturbation Hamiltonian in the upper plane for $\{\mib{s}_{k,a}\}$.
The same perturbation also appears in the lower plane for $\{\mib{s}_{k,b}\}$.
}
\label{fig:2}
\end{figure}

We consider a square lattice in the $xy$-plane and replace all of the edges with the diamond units, which leads to
\begin{equation}
   H_0=\sum_{\langle i,j \rangle}h_{i,j},
\end{equation}
where $\langle i,j \rangle$ represents a nearest-neighbor pair of the square lattice.
We make the direction of the bond spin-pairs parallel to the $z$-axis.
A schematic of $H_0$ is given in Fig.~\ref{fig:2}(a), 
and this lattice is called a diamond-like decorated square lattice.
The ground-state manifold of $H_0$ consists of the dimer covering states of the square lattice,
where a dimer means the tetramer ground state $\phi^g$.
For a square-lattice bond occupied by no dimer, the bond spin-pair is in the singlet state.
In contrary to the singlet dimer, which takes the central role in the ordinary RVB physics, 
our ``dimer'' has no orientation, which is easily observed from Eq.~(\ref{eq:2}), and our dimer coverings have orthogonality with the aid of bond spin-pairs.
These two properties of our manifold are usually assumed in the study of QDMs\cite{Moessner2008}.

%----------------------------------------------------
%----------------------------------------------------

In the geometry of our diamond-like decorated square lattice, we have three layers:
the center layer composed of edge spins at $z=0$,
the upper layer composed of $\{\mib{s}_{k,a}\}$ at $z>0$,
and the lower layer composed of $\{\mib{s}_{k,b}\}$ at $z<0$.
In this study, we investigate the effects of nearest-neighbor couplings within the upper and lower layers.
We write the perturbation Hamiltonian as
\begin{equation}
   H'=\delta\sum_{\langle k,k' \rangle}(\mib{s}_{k,a}\cdot\mib{s}_{k',a}+\mib{s}_{k,b}\cdot\mib{s}_{k',b}),
\end{equation}
where $\langle k,k' \rangle$ represents a nearest-neighbor pair of the square lattice composed of bond spin-pairs [see Fig.~\ref{fig:2}(b)].

\subsection{Matrix elements of a perturbation bond}%%%%%%%%

The perturbation operator between bond spin-pairs at sites $k$ and $k'$ is given by
\begin{equation}
   V_{k,k'}=\mib{s}_{k,a}\cdot\mib{s}_{k',a}+\mib{s}_{k,b}\cdot\mib{s}_{k'b}.
\end{equation}
When a linear term of spin operators operates on the singlet state $|s\rangle$, we obtain triplet states.
When the perturbation bond $V_{k,k'}$ operates on the state $|s\rangle_k|s\rangle_{k'}$, we have
\begin{equation}
   V_{k,k'}|s\rangle_k|s\rangle_{k'}=\frac{1}{2}\left(|t^0\rangle_k|t^0\rangle_{k'}-|t^+\rangle_k|t^-\rangle_{k'}-|t^-\rangle_k|t^+\rangle_{k'}\right),
\end{equation}
in which we note that both spin-pairs turn into triplet states.
For $|t^{\alpha}\rangle_k|s\rangle_{k'}$ ($\alpha=-,\;0$, or $+$), we have
\begin{equation}
   V_{k,k'}|t^{\alpha}\rangle_k|s\rangle_{k'}=\frac{1}{2}|s\rangle_k|t^{\alpha}\rangle_{k'},
\label{eq:ts}
\end{equation}
in which we note that the singlet and triplet sites are replaced by each other.
In general, a linear term of spin operators operating on a triplet state gives a linear combination of singlet and triplet states.
The reason for the lack of terms containing $|t\rangle_k|t\rangle_{k'}$ on the right-hand side of Eq.~(\ref{eq:ts}) is that
$V_{k,k'}$ is symmetric for $\mib{s}_{k,a}\leftrightarrow \mib{s}_{k,b}$ and $\mib{s}_{k',a}\leftrightarrow \mib{s}_{k',b}$.

Our lowest-energy manifold consists of all possible dimer coverings, in which a square-lattice site is prohibited from belonging to two or more dimers.
When $V_{k,k'}$ operates on a dimer covering, we obtain square-lattice sites belonging to two or more dimers.
Thus, we notice that our effective Hamiltonian has no first-order term and starts with the second-order term.

%%%%%%%%%%%%%%%%%%%%%%%%%%%%%%%%%%%%%%%%%%%%%%%%%%%%%%%%%%%%%%%%%%
\section{Second-Order Perturbation}\label{sec:3}
%%%%%%%%%%%%%%%%%%%%%%%%%%%%%%%%%%%%%%%%%%%%%%%%%%%%%%%%%%%%%%%%%%

\subsection{Effective Hamiltonian}%%%%%%%%%%%%%%%%%%%

We turn to the second-order effective Hamiltonian.
First, it should be noted that any possible second-order process cannot be created by perturbation bonds belonging to different plaquettes\cite{Comment}.
Because possible second-order processes are created by using the perturbation bonds on a plaquette twice,
the second-order effective Hamiltonian can be written in the following form:
\begin{equation}
   H_{\rm{eff}}=-t \hat{T}+\epsilon_2\hat{D}_2+\epsilon_1\hat{D}_1+\epsilon_0\hat{D}_0,
\label{eq:8}   
\end{equation}
where 
\begin{equation}
\begin{minipage}[c]{.80\textwidth}
\begin{center}
\includegraphics[width=.63\linewidth]{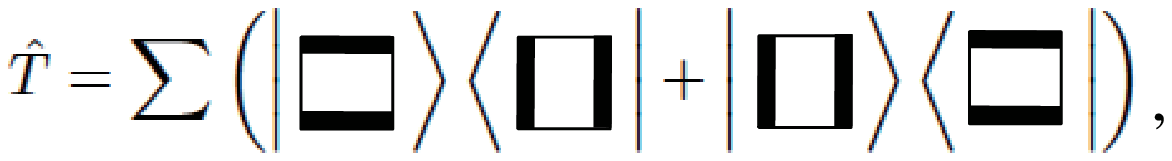}
\end{center}
\end{minipage}
\end{equation}
\begin{equation}
\begin{minipage}[c]{.90\textwidth}
\begin{center}
\includegraphics[width=.58\linewidth]{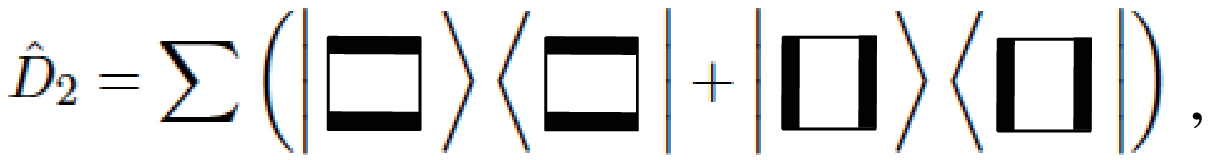}
\end{center}
\end{minipage}
\end{equation}
\begin{equation}
\begin{minipage}[c]{.9\textwidth}
\begin{center}
\includegraphics[width=.95\linewidth]{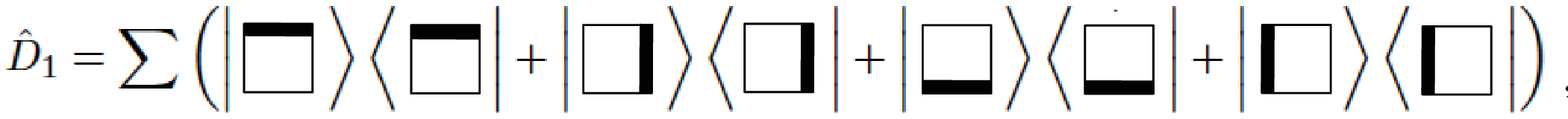}
\end{center}
\end{minipage}
\end{equation}
and
\begin{equation}
\begin{minipage}[c]{.85\textwidth}
\begin{center}
\includegraphics[width=.35\linewidth]{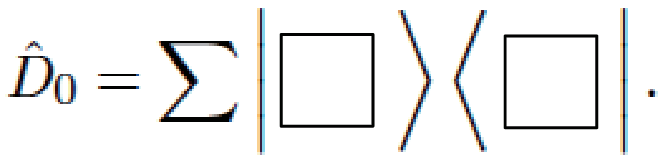}
\end{center}
\end{minipage}
\end{equation}
We can use the conditions
\begin{equation}
   \hat{D}_2+\hat{D}_1+\hat{D}_0=\sum 1=\mbox{[total number of plaquettes]}
\end{equation}
and
\begin{equation}
   \hat{D}_2+\frac{1}{2}\hat{D}_1=\mbox{[total number of dimers]}
\end{equation}
to eliminate $\hat{D}_1$ and $\hat{D}_0$ from $H_{\rm{eff}}$.
Omitting the constant terms, we have 
\begin{equation}
   H_{\rm{eff}}= -t \hat{T}+(\epsilon_2-2\epsilon_1+\epsilon_0)\hat{D}_2.
\end{equation}
Thus, we obtain an expression for the repulsion parameter $v$ as follows:
\begin{equation}
   v=\epsilon_2-2\epsilon_1+\epsilon_0.
\label{eq:v}   
\end{equation}
In the following, we perform explicit perturbation calculations for $t$, $\epsilon_2$, $\epsilon_1$, and $\epsilon_0$ and show that
\begin{equation}
   t=1.060\;\delta^2,\;\;\;v=0.
\label{eq:result}
\end{equation}
Equation (\ref{eq:result}) indicates that the present perturbation leads to the stabilization of a plaquette phase\cite{Leung1996}.

\subsection{Proof of $v=0$}%%%%%%%%%%%%%%%%%%%

%----------------------------------------------------
\begin{figure}[t]
\begin{center}
\includegraphics[width=.45\linewidth]{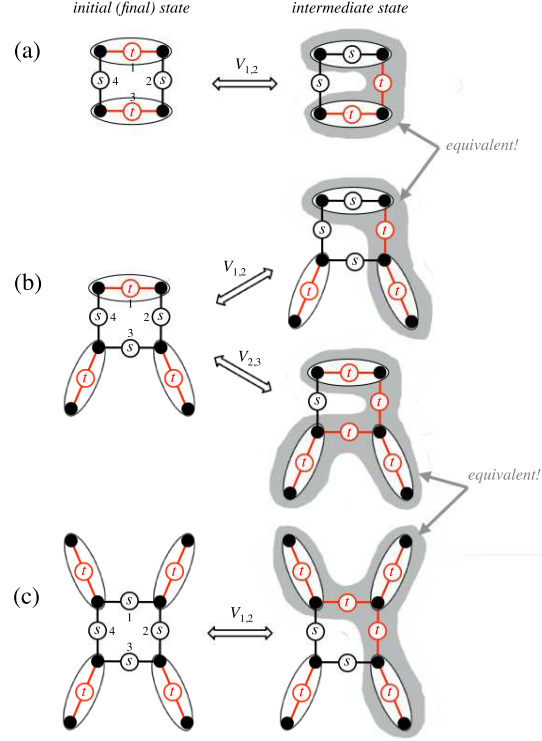}
\end{center}
\caption{(Color online) Perturbation processes for the diagonal terms for (a) $\epsilon_2$, (b) $\epsilon_1$, and (c) $\epsilon_0$.
The red bonds indicate that the bond spin-pair is in triplet states.
A red bond with an oval represents the tetramer ground state $\phi^g$.
A black bond with an oval represents the state that is obtained by replacing $|t\rangle$ in $\phi^g$ with $|s\rangle$.
}
\label{fig:3}
\end{figure}
%----------------------------------------------------

Here, we consider the diagonal terms in the effective Hamiltonian.
Concentrating on the plaquette to which the perturbation bonds producing a process belong,
we show the possible second-order processes in Fig.~\ref{fig:3}. 
In Fig.~\ref{fig:3}(a), the initial state has two dimers on the plaquette.
The process shown in this figure is that produced by $V_{1,2}$.
In addition, $V_{2,3}$, $V_{3,4}$, and $V_{4,1}$ also essentially result in the same process, 
and the sum of these four yields $\epsilon_2$ is defined in Eq.~(\ref{eq:8}).
The initial state in Fig.~\ref{fig:3}(c) has no dimer on the plaquette.
In addition to the process produced by $V_{1,2}$, we also have the same contributions from those produced by $V_{2,3}$, $V_{3,4}$, and $V_{4,1}$.
The parameter $\epsilon_0$, which is defined in Eq.~(\ref{eq:8}), is determined by the sum of these four process.
The initial state in Fig.~\ref{fig:3}(b) has one dimer on the plaquette.
In this case, we have two types of process: 
the first one is produced by $V_{1,2}$ and $V_{1,4}$, and the second one is produced by $V_{2,3}$ and $V_{3,4}$.
Noting that practical calculations can be carried out in the gray-colored clusters in Fig.~\ref{fig:3},
we find that the first (second) process in Fig.~\ref{fig:3}(b) gives the same result as the process in Fig.~\ref{fig:3}(a) [Fig.~\ref{fig:3}(c)].
Taking the total number of equivalent processes into consideration, we obtain
\begin{equation}
   \epsilon_1=\frac{1}{2}(\epsilon_0+\epsilon_2).
\end{equation}
Substituting this relation into Eq.~(\ref{eq:v}), we conclude that 
\begin{equation}
   v=0.
\end{equation}

\subsection{Calculation process for the hopping parameter $t$}%%%%%%%%%%%%%%%%%%%

\begin{figure}[t]
\begin{center}
\includegraphics[width=.55\linewidth]{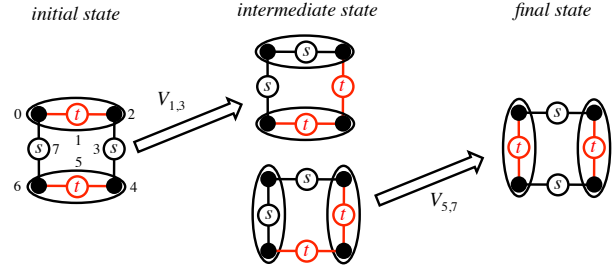}
\end{center}
\caption{(Color online) Second-order perturbation process for dimer pair hopping.
Two intermediate states, $V_{1,3}|\mbox{initial state}\rangle$ and $V_{5,7}|\mbox{final state}\rangle$, are shown.
Note that these two states are not linearly independent. 
Moreover, these two states are not the eigenstates of $H_0$;
therefore, we have to expand them by using the eigenstates of $H_0$, which is carried out numerically.}
\label{fig:4}
\end{figure}

We turn to the calculation of the off-diagonal terms.
A possible process is shown in Fig.~\ref{fig:4}.
Defining the site indices as shown in this figure with $V_{1,3}$ operating on the initial state $|s_{3};\phi^g_{0,2,1};\phi^g_{4,6,5}\rangle$, we have
\begin{eqnarray}
   V_{1,3}|\phi^g_{0,2,1};s_{3}\rangle&&\hspace{-6mm}
   \nonumber\\
   &&\hspace{-25mm}=
   \frac{1}{2\sqrt{3}}|s\rangle_{1}
   \left[
   |\!\downarrow\downarrow\rangle_{0,2}|t^+\rangle_{3}-
   \frac{|\!\uparrow\downarrow\rangle_{0,2}+|\downarrow\uparrow\rangle_{0,2}}{\sqrt{2}}|t^0\rangle_{3}+
   |\!\uparrow\uparrow\rangle_{0,2}|t^-\rangle_{3}
   \right].
\end{eqnarray}
%----------------------------------------------------
%----------------------------------------------------
If the bond spin-pairs at sites $3$ and $5$ are in triplet states, then a connected cluster $(2,3,4,5,6)$ appears in the unperturbed system.
For the unperturbed Hamiltonian $h_{2,4}+h_{4,6}$ for this cluster, 
we write the eigenvalues and eigenstates as $E_n+2\lambda$ and $|\Psi^n\rangle_{2,3,4,5,6}$, respectively.
Denoting the intermediate states as $|\sigma_0;s_{1};\Psi^n_{2,3,4,5,6}\rangle$ with $\sigma=\uparrow,\downarrow$, we have
\begin{eqnarray}
   \langle\sigma_0;s_{1};\Psi^n_{2,3,4,5,6}|V_{1,3}|s_{3};\phi^g_{0,2,1};\phi^g_{4,6,5}\rangle
   &&\hspace{-6mm}
   \nonumber\\
   &&\hspace{-50mm}
   =\delta_{\sigma,\downarrow}\langle\Psi^n_{2,3,4,5,6}|\varphi_{2,3};\phi^g_{4,6,5}\rangle+
   \delta_{\sigma,\uparrow}\langle\Psi^n_{2,3,4,5,6}|
   \bar{\varphi}_{2,3};\phi^g_{4,6,5}\rangle,
\label{eq:21}   
\end{eqnarray}
where
\begin{equation}
   |\varphi_{2,3}\rangle=\frac{1}{2\sqrt{3}}
   \left(
   |\!\downarrow_2;t^+_{3}\rangle-\frac{|\!\uparrow_2;t^0_{3}\rangle}{\sqrt{2}}
   \right),\;\;
   |\bar{\varphi}_{2,3}\rangle=\frac{1}{2\sqrt{3}}
   \left(
   |\!\uparrow_2;t^-_{3}\rangle-\frac{|\!\downarrow_2;t^0_{3}\rangle}{\sqrt{2}}
   \right).
\end{equation}
The energy denominator of the intermediate states is given by
\begin{equation}
   E_n+2\lambda-2(\lambda-2)=E_n+4.
\end{equation}
Next, when $V_{5,7}$ operates on the final state $|s_{5};\phi^g_{0,6,7};\phi^g_{2,4,3}\rangle$, we have
\begin{eqnarray}
   V_{5,7}|\phi^g_{0,6,7};s_{5}\rangle&&\hspace{-6mm}
   \nonumber\\
   &&\hspace{-25mm}
   =
   \frac{1}{2\sqrt{3}}|s\rangle_{7}
   \left[
   |\!\downarrow\downarrow\rangle_{0,6}|t^+\rangle_{5}-
   \frac{|\!\uparrow\downarrow\rangle_{0,6}+|\downarrow\uparrow\rangle_{0,6}}{\sqrt{2}}|t^0\rangle_{5}+
   |\!\uparrow\uparrow\rangle_{0,6}|t^-\rangle_{5}
   \right].
\end{eqnarray}
Thus, the matrix element between the final state and an intermediate state 
$|\sigma_0;s_{7};\Psi^n_{2,3,4,5,6}\rangle$ ($\sigma=\uparrow$ or $\downarrow$) is given by
\begin{eqnarray}
   \langle\sigma_0;s_{7};\Psi^n_{2,3,4,5,6}|V_{5,7}|s_{5};\phi^g_{0,6,7};\phi^g_{2,4,3}\rangle&&\hspace{-6mm}
   \nonumber\\
   &&\hspace{-50mm}=
   \delta_{\sigma,\downarrow}\langle\Psi^n_{2,3,4,5,6}|\varphi_{6,5};\phi^g_{2,4,3}\rangle+
   \delta_{\sigma,\uparrow}\langle\Psi^n_{2,3,4,5,6}|
   \bar{\varphi}_{6,5};\phi^g_{2,4,3}\rangle.
\label{eq:29}
\end{eqnarray}
Using Eqs. (\ref{eq:21}) and (\ref{eq:29}), we obtain
\begin{eqnarray}
   t&&\hspace{-6mm}=
   4\delta^2\sum_{\sigma=\uparrow,\downarrow}\sum_n
   \langle s_{5};\phi^g_{0,6,7};\phi^g_{2,4,3};s_{1}|V_{5,7}|\sigma_0;s_{1};s_{7};\Psi^n_{2,3,4,5,6}\rangle
   \frac{1}{E_n+4}
   \nonumber\\&&\hspace{40mm}
   \times\langle\sigma_0;s_{1};s_{7};\Psi^n_{2,3,4,5,6}|V_{1,3}|s_{3};\phi^g_{0,2,1};\phi^g_{4,6,5};s_{7}\rangle
   \nonumber\\&&\hspace{-6mm}=
   8\delta^2\sum_n\frac{\langle\varphi_{6,5};\phi^g_{2,4,3}|\Psi^n_{2,3,4,5,6}\rangle\langle\Psi^n_{2,3,4,5,6}|\varphi_{2,3};\phi^g_{4,6,5}\rangle
   }{E_n+4}.
\end{eqnarray}
We numerically calculate $\{E_n, \Psi^n\}$ for the subspace with a total $S^z$ of $1/2$, whose total number is 18,
and obtain
\begin{equation}
   t/\delta^2=1.06026786.
\end{equation}

%%%%%%%%%%%%%%%%%%%%%%%%%%%%%%%%%%%%%%%%%%%%%%%%%%%%%%%%%%%%%%%%%%
\section{Summary}\label{sec:4}
%%%%%%%%%%%%%%%%%%%%%%%%%%%%%%%%%%%%%%%%%%%%%%%%%%%%%%%%%%%%%%%%%%

We have derived a QDM as the low-energy effective Hamiltonian of a diamond-like decorated square-lattice Heisenberg antiferromagnet.
The QDM has a finite hopping parameter and no interaction, which results in the plaquette phase.

In our diamond-like decorated square lattice, the direction of bond spin-pairs is orthogonal to the plane formed by edge spins.
We can consider other geometries. 
If we choose the direction of bond spin-pairs to be parallel to the edge-spin plane, as in Fig. 1 of Ref. 9,
then the further neighbor coupling is not symmetric with respect to $\mib{s}_{k,a}\leftrightarrow\mib{s}_{k,b}$.
As a result, the matrix element of a perturbation bond becomes more complicated.
Then,  an interaction term may appear, which depends on the coupling parameter $\lambda$ within a bond spin-pair.
It will be interesting to study the ground-state phase diagram as a function of $\lambda$.
From the viewpoint of the exploration of RVB liquids, it will also be interesting to construct QDMs on nonbipartite and simple cubic lattices.
Since our model contains only simple Heisenberg-type couplings, QDMs can be realized in a quantum simulator by optical lattices\cite{Bloch2008}.
Thus, our construction of QDMs from Heisenberg models provides a clear path to realizing QDMs in a quantum simulator, which may lead to new progress in RVB physics.


\begin{thebibliography}{99}
\bibitem{Anderson1973} P.~W.~Anderson, Math. Res. Bull. \textbf{8}, 153 (1973).
\bibitem{Rokhsar1988} D. S. Rokhsar and S. A. Kivelson, Phys. Rev. Lett. \textbf{61}, 2376 (1988).
\bibitem{Leung1996} P. W. Leung, K. C. Chiu, and K. J. Runge, Phys. Rev. B \textbf{54}, 12938 (1996).
\bibitem{Moessner2008} R. Moessner and K. S. Raman, arXiv:cond-mat/0809.3051v1.
\bibitem{Yan2011} S. Yan, D. A. Huse, and S. R. White, Science \textbf{332}, 1173 (2011).
\bibitem{Depenbrock2012} S. Depenbrock, I. P. McCulloch, and U. Schollw\"ock, Phys. Rev. Lett. \textbf{109}, 067201 (2012).
\bibitem{Fujimoto} S. Fujimoto, Phys. Rev. B \textbf{72}, 024429 (2005).
\bibitem{Morita2016} K. Morita and N. Shibata, J. Phys. Soc. Jpn. \textbf{85}, 033705 (2016).
\bibitem{Bloch2008} I. Bloch, J. Dalibard, and W. Zwerger, Rev. Mod. Phys.  \textbf{80}, 885 (2008).
\bibitem{Takano1996} K. Takano, K. Kubo, and H. Sakamoto: J. Phys.: Condens. Matter \textbf{8}, 6405 (1996).
\bibitem{Canova2010} L. $\check{\rm C}$anov$\acute{\rm a}$ and J. Stre$\check{\rm c}$ka, Phys. Status Solidi B \textbf{247}, 433 (2010).
\bibitem{Comment} When a perturbation bond operates on a dimer covering state, 
two edge spins on a diagonal position of the plaquette become ``defects,''
where a defect means an edge spin that belongs to two or more dimers or to no dimer. 
Any perturbation bond on another plaquette cannot remove these two defects.
\end{thebibliography}
\end{document}